\documentclass[11pt, a4paper]{article}
\usepackage[utf8]{inputenc}
\usepackage[T1]{fontenc}
\usepackage{amsmath, amssymb, amsfonts} 
\usepackage{graphicx}                   
\usepackage{booktabs}                   
\usepackage{xcolor}                     
\usepackage{listings}                   
\usepackage{hyperref}                   

\usepackage[numbers,sort&compress]{natbib}
\usepackage{dcolumn}
\usepackage{bm}

\usepackage{listings}
\usepackage{xcolor}

\begin{document}


\title{Joint Optimization of Neural Autoregressors via Scoring rules}

\author{Jonas Landsgesell}

\date{\today}

\maketitle

\section{Abstract}
Non-parametric distributional regression has achieved significant milestones in recent years. Among these, the Tabular Prior-Data Fitted Network (TabPFN) \cite{hollmann2022tabpfn} has demonstrated state-of-the-art performance on various benchmarks. However, a persistent challenge remains in extending these grid-based approaches to a truly multivariate setting.In a naive non-parametric discretization with $N$ bins per dimension, the complexity of an explicit joint grid scales as $\mathcal{O}(N^d)$. This exponential growth—the "curse of dimensionality"—renders multivariate grid prediction computationally prohibitive even for low-dimensional outputs. Beyond the memory bottleneck, this scaling is particularly detrimental in low-data regimes, as the final projection layer would require $hidden\_dim \times N^d$ parameters, leading to severe overfitting and intractability.

\section{\label{sec:level1} Introduction}

Traditional regression models primarily focus on predicting a single point estimate (e.g., the mean or median) of a target variable, given a set of input features. While useful for many applications, this approach often falls short when a more complete understanding of the underlying data-generating process is required. Distributional Regression, in contrast, aims to model the entire conditional distribution of the target variable(s) given the inputs\cite{kneib2023rage}\cite{gneiting2014probabilistic}\cite{gneiting2008probabilistic}. Instead of merely predicting $E[Y|X]$, it seeks to estimate $P(Y|X)$. 
The shift from point-wise estimates to full characterization of the conditional density $P(\mathbf{y}|\mathbf{x})$ is fundamentally rooted in the seminal framework of probabilistic forecasting established by Gneiting et al. \cite{gneiting2007strictly}. Their work provides the rigorous mathematical justification for using strictly proper scoring rules, ensuring that the model is incentivized to achieve both 'sharpness' and 'calibration'—a departure from traditional regression that often ignores the higher-order moments of the residual distribution.
Therefore distributional regression provides a richer and more comprehensive output, quantifying not just the central tendency but also the spread, skewness, and other moments of the conditional distribution. This is particularly valuable in fields like risk management, uncertainty quantification in scientific simulations, or personalized medicine, where understanding the full range of possible outcomes and their probabilities is crucial for informed decision-making\cite{gneiting2011making}. 
The question about how good different probabilistic forecasts are, can be answered by suitable proper scoring rules\cite{gneiting2007strictly} (most probably use-case dependent, because each scoring-rule has its own trade-off).

\subsection{The Curse of Dimensionality in Naive Multivariate Approaches}

Extending distributional regression to multiple output dimensions introduces a fundamental scaling challenge known as the curse of dimensionality. A naive approach to modeling a joint distribution involves discretizing the $D$-dimensional output space into a hyper-grid and predicting a probability density for every resulting cell.If each dimension is discretized into $B$ bins, the total number of output parameters required to describe the joint distribution grows as $B^D$. This exponential scaling quickly becomes intractable; for even a moderate number of dimensions and a standard bin resolution, the parameter space exceeds the memory capacity of modern hardware.The bottleneck is most apparent in the final projection layer of a neural network. A feed-forward layer mapping a hidden representation to this grid would require a weight matrix of size $hidden\_dim \times B^D$. Such an architecture suffers from several critical flaws 
\begin{itemize}
    \item VRAM Constraints: The memory footprint for storing the weights alone can easily exceed the limits of current GPU clusters.
    \item Data Scarcity: Accurately estimating values in a $B^D$ space requires an exponentially large dataset to avoid the sparse-data problem, where most grid cells contain no observations.
    \item Computational Latency: The overhead for computing and normalizing (via softmax) an exponential number of logits creates severe bottlenecks during both training and inference.
\end{itemize}
These constraints make direct, explicit grid-based modeling of joint distributions expensive. This necessitates a more efficient parameterization that captures the inter-variable dependencies without the overhead of an exhaustive grid.

\section{\label{sec:level2}Proposed Solution}
We propose to lower the amount of needed parameters/compute time/data by replacing the traditional feed forward vector for computing the probability vector with a rnn or masked transformer if the random variables for the outcome are dependent or low rank matrices if the random variables are independent.

Our proposed simple architecture allows to compute probabilitstic predictions in more than one dimension (multi output regression), which is why we called it
Joint Optimization of Neural Autoregressors via Scoring rules (`JonasNet`) architecture. We also propose to use well known proper scoring rules like the energy-score (higher dimensional CRPS) or the variogramm based proper scoring rule. The architecture contains a

\begin{itemize}
    \item \emph{Feature Extractor:} A multi-layer perceptron (MLP) that processes the input features `X` and transforms them into a rich, lower-dimensional representation (latent context vector).
    \item \emph{Autoregressive Decoder (RNN/Transformer):}
    \begin{itemize}
        \item RNN-based approach: An RNN (e.g., GRU or LSTM) takes the latent context vector from the feature extractor as its initial hidden state. For each output dimension `d` (from `1` to `D`), the RNN predicts the conditional distribution of $Y_d$, using the actual (during training via teacher forcing) or predicted (during inference) value of $Y_{d-1}$ as input for the next step. This sequential processing naturally enforces the conditional dependencies.
        \item Causally Masked Transformer approach: A transformer encoder/decoder architecture where the input to the decoder consists of the context vector $X$ and an embedding of the previously generated $Y$ values. A causal (or look-ahead) mask is applied to the self-attention mechanism within the transformer, ensuring that the prediction for $Y_d$ only depends on $X$ and $Y_1, ..., Y_{d-1}$, but not on $Y_{d+1}, ..., Y_D$. This allows for parallel computation during training while maintaining the autoregressive property for inference.
    \end{itemize}
    \item Bin Head: For each predicted output dimension, a small MLP (the 'bin head') takes the hidden state of the RNN or the transformer's output for that dimension and outputs logits over $n_bins$ for a discretized probability distribution. These logits are then converted into probabilities (e.g., using softmax) representing the likelihood of the output falling into specific intervals.
\end{itemize}
For independent random variables, a lightweight solution is to replace the above recurrent network/transformer layers, with low rank matrices with enough capacity to learn the patterns of interest. 

\subsection*{Methodological Comparison: Multivariate Density Estimation}

The sequential nature of the RNN or the causal masking of the transformer is specifically designed to capture inter-dimensional dependencies. When predicting $P(Y_d | Y_1, ..., Y_{d-1}, X)$, the model implicitly learns how $Y_d$ is influenced by the preceding `Y` values. For instance, in an RNN, the hidden state at step `d` encodes information about `X` and $Y_1, ..., Y_{d-1}$. This hidden state then informs the prediction of $Y_d$. Similarly, in a causally masked transformer, the attention mechanism allows each $Y_d$ prediction to attend to all preceding `Y`s and the initial `X` representation, effectively integrating their influence. This allows the model to learn complex, non-linear relationships between the output dimensions, moving beyond simple mean predictions to capture how the *distribution* of one output variable changes based on the values of others.

\subsubsection*{Autoregressive Coupling (RNN/Transformer)}
\begin{figure*}
  \centering
  \includegraphics[width=0.4\textwidth]{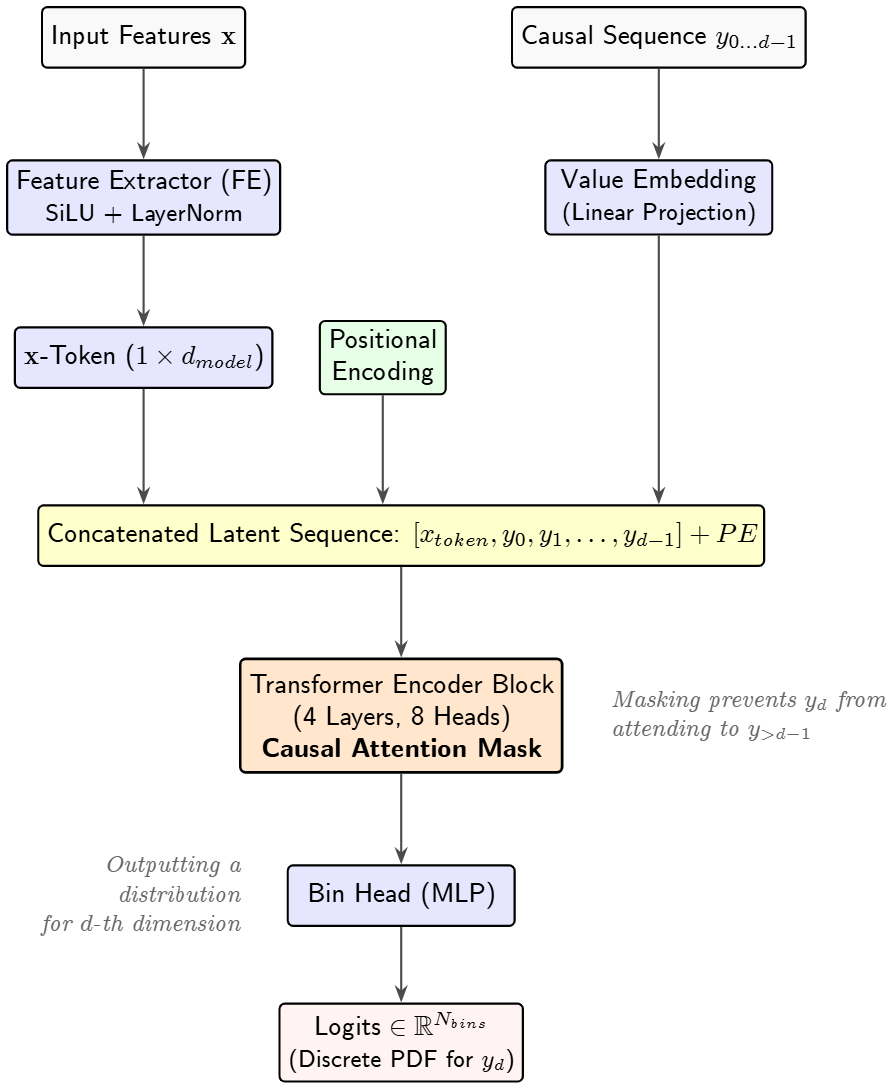} 
  \caption{\label{fig:architecture}Architecture of JonasNet.}
\end{figure*}
For systems exhibiting strong inter-variable dependencies, the distribution is factorized using the chain rule:
\begin{equation}
    P(y_1, y_2, \dots, y_d \mid \mathbf{x}) = \prod_{i=1}^{d} P(y_i \mid y_{<i}, \mathbf{x})
\end{equation}
The hidden state of an RNN or the attention mechanism of a Transformer enables the modeling of the \textit{joint distribution}. This is strictly required when physical coupling exists between the residuals of $y_i$, ensuring that generated samples remain consistent with physical constraints.

\subsubsection*{Marginal Coupling (Low-Rank/LoRA)}
If the target variables $y_1, \dots, y_d$ are conditionally independent given $\mathbf{x}$, the problem reduces to estimating the marginal densities:
\begin{equation}
    P(y_1, y_2, \dots, y_d \mid \mathbf{x}) \approx \prod_{i=1}^{d} P(y_i \mid \mathbf{x})
\end{equation}
The low-rank projection ($\text{LoRA}$) provides an efficient representation of the shared feature space but neglects the stochastic coupling of the outputs during the sampling process.

\section{Methodology}

\subsection{Scoring Rules and Probabilistic Forecasting}
A central challenge in multivariate forecasting is the transition from point estimates to full distribution characterization. Point estimates  (like the median from mean absolute minimization or the conditional mean from mean squared error minimization) are inherently brittle, particularly in the presence of epistemic uncertainty or multimodal data-generating processes. By predicting a binned probability density function—essentially a probability mass function (PMF) over a discrete grid—we can leverage strictly proper scoring rules. This grid can have equal or unequal bin widths.

As established by Gneiting and Raftery (2007)\cite{gneiting2007strictly}, a scoring rule $S(\vec{y}, \hat{\vec{P}})$ is strictly proper if the expected score $E_{\vec{y}\sim \vec{Y}}[S(\vec{y}, \vec{p})]$ is minimized if and only if the forecast $\hat{\vec{p}}$ matches the true distribution $\vec{Y}\sim F$. For our discretized output space, we can evaluate the local logarithmic score (Cross-Entropy), the \textbf{Energy Score}, the latter being the multivariate generalization of the Continuous Ranked Probability Score (CRPS) or other scoring rules for multivariate settings like the variogramm-based scoring rule
\begin{equation}
S_{\gamma}( \vec y,\hat{\vec p})
=
\sum_{1 \le i < j \le d}
w_{ij}
\left(
\lvert y_i - y_j \rvert^{\gamma}
-
\lvert \hat{p}_i - \hat{p}_j \rvert^{\gamma}
\right)^2 ,
\end{equation}
where $\vec y \in \mathbb{R}^d$ denotes the observed outcome vector,
$\hat{\vec p} \in \mathbb{R}^d$ the predicted probability (mean) vector,
$w_{ij} \ge 0$ pairwise weights, and
$\gamma \in (0,2]$ the variogram order.

\subsection{Discrete Energy Score on a Grid}
To bypass the limitations of parametric assumptions, we directly estimate the non-parametric likelihood. We define a grid where each bin $i$ is associated with a position vector $\vec{r}_{m,i}$. For a $d$-dimensional target vector $\vec{y}$, the Energy Score\cite{gneiting2007strictly} $S(\vec{y},\hat{\vec{p}})$ for a predicted probability vector $\vec{p}$ and an observation $\vec{y}$ is defined as:

\begin{equation}
    S(\vec{y},\hat{\vec{p}}) = \sum_{i=1}^{N} p_i \|\vec{r}_{m,i} - \vec{y}\|^\beta - \frac{1}{2} \sum_{i,j=1}^{N} p_i p_j \|\vec{r}_{m,i} - \vec{r}_{m,j}\|^\beta
\end{equation}

where the sum runs over all bin indices of the high dimensional grid, $\beta \in (0, 2)$ is a divergence exponent (we chose $\beta=1$). The first term represents the expected $L^p$-distance between the predicted distribution and the realized observation $\vec{y}$. The second term serves as a kernel-based regularization, accounting for the geometry of the bin positions $\vec{r}_{m,i}$. This ensures the loss is sensitive to the spatial structure of the discretization; mass placed in a bin far from $\vec{y}$ is penalized more heavily than mass placed in an adjacent bin.

\subsection{Overcoming the Dimensionality Constraint}
While grid-based methods traditionally suffer from the curse of dimensionality—rendering a $50^{7}$ grid computationally intractable if a feed forward network is used for projection (due to memory limitations)—our architecture models the joint distribution $P(\vec{y}|x)$ as a sequence of conditional discretized probabilities. This approach allows us to:
\begin{itemize}
    \item Apply grid-based loss functions like the Energy Score across multi-target settings.
    \item Trivially deal with multi-target settings where we want to predict $P(\vec{y}|x)$ without requiring explicit covariance matrix inversion or Gaussian assumptions.
    \item Model multimodal generative outcomes by sampling from the discretized probability mass function.
\end{itemize}

To-date a non-parametric probability prediction in higher dimensions is not mainstream in machine-learning or data science.
We suggest that non-parametric multivariate density estimation will become a standard primitive in machine learning as computational efficiency bottlenecks—such as the curse of dimensionality—are addressed by autoregressive architectures like the one proposed here.

\section{Results}

When there is correlation Cov($y_i,y_j$) between target variables, our model picks up on this and outperforms univariately trained xgboost.

In multivariate regression, modeling the joint distribution allows the network to act as a shrinkage estimator, which reduces total risk by "pooling" information across targets and pulling individual noisy predictions toward the learned conditional manifold—thereby outperforming independent univariate models somewhat reminding of the Stein Paradox for three or more dimensions (even for independent random variables).

\section*{Data Generation Process}

The synthetic dataset is generated using a latent variable model where two underlying signals are coupled and transformed into an observation space. For a given input $x \in [0, 10]$, the process is defined as follows:

\subsection*{1. Latent Source Space}
The ground truth signals $\mathbf{s}(x) = [s_1(x), s_2(x)]^T$ represent a non-linear relationship:
\begin{align}
    s_1(x) &= \sin(x) \\
    s_2(x) &= \frac{1}{2} s_1(x)^2
\end{align}

\subsection*{2. Stochastic Component}
To simulate realistic measurement conditions, we introduce additive Gaussian noise $\boldsymbol{\epsilon}(x)$ with heteroscedastic properties in the first dimension:
\begin{equation}
    \boldsymbol{\epsilon}(x) \sim \mathcal{N}\left(\mathbf{0}, \boldsymbol{\Sigma}(x)\right), \quad 
    \boldsymbol{\Sigma}(x) = \begin{pmatrix} 
    (0.1 + 0.05x)^2 & 0 \\ 
    0 & 0.1^2 
    \end{pmatrix}
\end{equation}

\subsection*{3. Observation Space Transformation}
The final observations $\mathbf{y}(x)$ are obtained by applying a rotation matrix $\mathbf{R}(\theta)$ to the noisy source signals, inducing a coupling between the observed dimensions:
\begin{equation}
    \mathbf{y}(x) = \mathbf{R}(\theta) \left( \mathbf{s}(x) + \boldsymbol{\epsilon}(x) \right)
\end{equation}
where $\theta = 30^\circ$ and the rotation matrix is defined as:
\begin{equation}
    \mathbf{R}(\theta) = \begin{pmatrix} 
    \cos\theta & -\sin\theta \\ 
    \sin\theta & \cos\theta 
    \end{pmatrix}
\end{equation}

The validation of the model performance (MSE) is conducted with hold out data against the rotated ground truth $\mathbf{y}_{GT}(x) = \mathbf{R}(\theta)\mathbf{s}(x)$.
\begin{table}[htbp]
\centering
\caption{Comparison of Mean Squared Error (MSE) between JonasNet and XGBoost on held out data for the described toy dataset with a total sample size of 250 and 80-20 train-test split }
\label{tab:model_comparison}
\begin{tabular}{lrrr}
\toprule
\textbf{Model} & \textbf{Total MSE} & \textbf{MSE1} & \textbf{MSE2} \\ \midrule
JonasNet       & \textbf{0.00702}            & \textbf{0.00535}            & \textbf{0.00870}            \\
XGBoost        & 0.04855            & 0.06302            & 0.03408            \\ \bottomrule
\end{tabular}
\end{table}

\begin{figure*}
  \centering
  \includegraphics[width=0.7\textwidth]{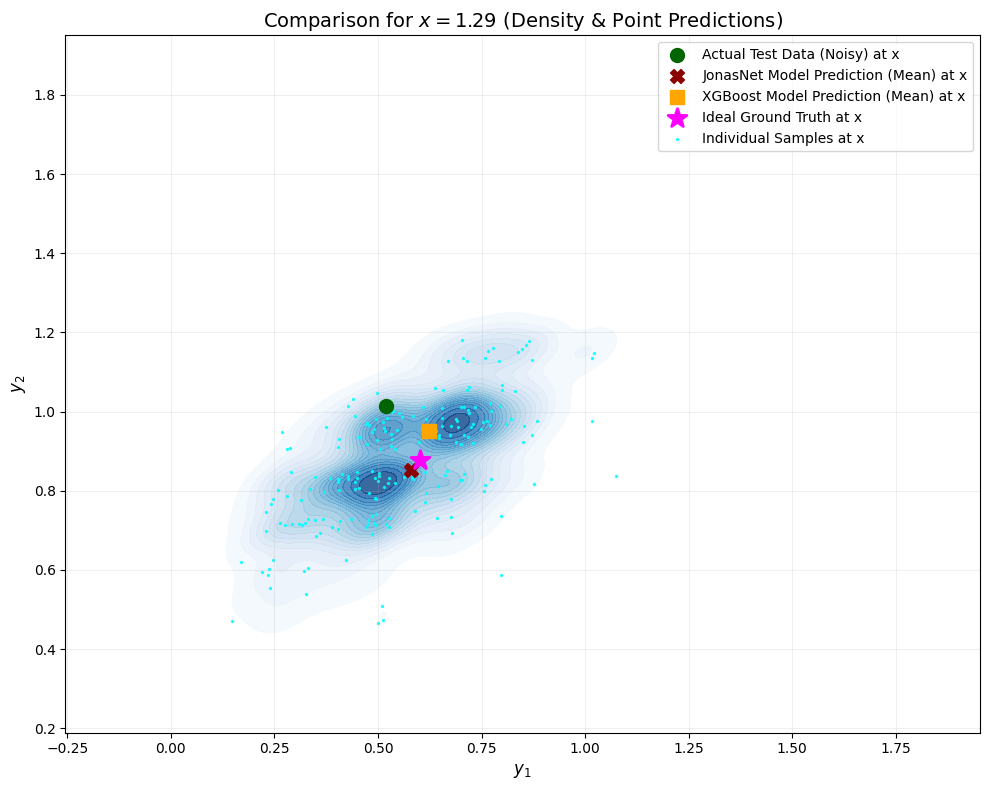} 
  \caption{\label{fig:res}\textbf{Posterior Predictive Density at $x=1.29$}. The blue contours represent the JonasNet predicted distribution $P(\mathbf{y}|x)$, capturing the heteroscedastic coupling. While the actual noisy observation (green circle) is displaced by stochastic fluctuations, the JonasNet mean (red X) provides a superior approximation of the ideal ground truth (magenta star) compared to the XGBoost baseline with two independently fitted models per dimension(orange square).}
\end{figure*}

\section{Inference}
After training e can compute a discretized approximation of any functional of the probability density function through the use of the gridded probabability mass function.
This involves any quantile function, mean estimators, median estimators, variance estimators, kurtosis estimators and so on.

\subsection{Quantifying Uncertainty}

A key advantage of distributional regression is its ability to quantify uncertainty, and our proposed architecture fully embraces this. By predicting a probability distribution (over bins) for each conditional output, `JonasNet` inherently provides a measure of prediction uncertainty. For any given input `X`:
\begin{itemize}
    \item Point Predictions: The expected value (mean) or median, Variance, ... can be calculated from the predicted probability distribution `P(Y|X)` for each dimension.
    \item Uncertainty Intervals: Credible intervals (e.g., 90\% confidence intervals) can be directly derived from the cumulative distribution function (CDF) inferred from the predicted binned probabilities. This allows us to state not just what `Y` is likely to be, but also the range within which it is expected to fall with a certain probability. Coverage and Interval Score should be checked on held out data (repeated cross-validation) prior to decision making to assess the quality of the PMF-derived intervals.
    \item \dots
\end{itemize}

\subsection{Drawing samples from the estimated PMF}
The model's output provides a discrete probability mass function (PMF) over $N$ bins for each target dimension
\begin{equation}
PMF(Y=\vec{y}|\vec{X}=\vec{x})=pdf(Y=\vec{y}|\vec{X}=\vec{x}) \Delta^D.
\end{equation}
To transform these discrete logits into continuous realizations $\hat{\mathbf{y}}$, we implement a stochastic sampling routine:

\begin{enumerate}
    \item \textbf{Feature Expansion:} The input $\mathbf{x}$ is replicated $n$ times to allow for parallel Monte Carlo sampling of the output space.
    \item \textbf{Multinomial Draw:} For each dimension $j \in \{1, \dots, d\}$, a bin index $k$ is sampled based on the predicted probabilities:
    \begin{equation}
        k_j \sim \text{Multinomial}(p_{j,1}, \dots, p_{j,N})
    \end{equation}
    \item \textbf{Intra-bin Interpolation:} To avoid quantization artifacts and ensure a continuous support, the final scaled sample $s_j$ is drawn uniformly from within the identified bin intervals:
    \begin{equation}
        s_j = B_{k_j} + \epsilon \cdot (B_{k_j+1} - B_{k_j}), \quad \epsilon \sim \mathcal{U}(0,1)
    \end{equation}
    where $B$ denotes the predefined bin edges.
\end{enumerate}
Finally, the samples are mapped back to the original physical units using the inverse transform of the scaling statistics: $\hat{\mathbf{y}} = \text{SC}^{-1}(\mathbf{s})$.

These samples can then be used for Monte Carlo simulations, sensitivity analysis, or to visualize the uncertainty in the multivariate output space (e.g., via 2D kernel density estimates for pairs of output dimensions). This comprehensive uncertainty quantification is vital for robust decision-making in real-world scenarios where irreducible data noise and model uncertainty are prevalent.

\section{Contribution}
In this work, we introduce a solution for multivariate non-parametric distributional regression. Our contributions are summarized as follows:

\begin{itemize}
    \item \textbf{Autoregressive Joint Modeling:} We propose and evaluate a novel architecture that factorizes the joint distribution $P(\mathbf{y}|\mathbf{x})$ into a sequence of conditional discretized densities using RNNs and causally masked Transformers. This effectively bypasses the $\mathcal{O}(N^d)$ "curse of dimensionality" inherent in explicit grid-based multivariate modeling.
    \item \textbf{Low-Rank Independent Parameterization:} For scenarios with conditionally independent targets, we introduce a LoRA-style (Low-Rank Adaptation) projection head. This allows for shared feature extraction while maintaining a lightweight parameter footprint compared to standard multi-head MLP outputs.
    \item \textbf{Direct PMF Sampling \& De-quantization:} We implement a stochastic inference routine using multinomial sampling across the predicted probability mass functions (PMFs), combined with intra-bin uniform interpolation to generate continuous, non-quantized realizations for Monte Carlo downstream tasks.
    \item \textbf{Proper Scoring Rule Optimization:} We integrate strictly proper scoring rules, specifically the multivariate Energy Score and Variogram-based scores, into the training of discretized neural autoregressors. This ensures the model is incentivized toward both sharpness and calibration without parametric assumptions.
    \item \textbf{Empirical Validation of Neural Shrinkage:} We demonstrate that joint optimization can allow the model to act as a shrinkage estimator, capturing inter-variable correlations (stochastic coupling) that independent univariate models fail to resolve.
\end{itemize}

\section{Outlook}
We plan to integrate the proposed architecture which helps reducing impact of the curse of dimensionality into a Tabular Foundation Model, pretrained on high-dimensional synthetic data. By generating diverse priors from coupled stochastic processes and physical simulations, the model can learn to internalize the "grammar" of multivariate dependencies before encountering real-world tasks. We anticipate that this joint optimization will induce a shrinkage effect providing additional noise reduction and predictive stability compared to independent univariate models.

\section{Appendix}

The pseudo code for reproducing the proposed causally masked transformer encoder architecture is 

\begin{lstlisting}
# Architecture: Transformer-based Autoregressive Binner
Procedure JonasNet_Transformer(x_input, midpoints, y_targets):
    # 1. Feature Extraction
    x_token = SiLU(LayerNorm(Linear(x_input))) 
    
    # 2. Sequence Preparation
    If Training:
        # Teacher Forcing: Shift targets and prepend zero
        y_seq = Concat([ZeroToken, y_targets[..., :-1]])
    Else:
        # Autoregressive Loop
        y_seq = Initialize_with_Zeros(batch, output_dim)
    
    # 3. Embedding & Context
    embeddings = Linear_Embedding(y_seq) + Positional_Encoding
    full_sequence = Concat([x_token, embeddings])
    
    # 4. Causal Attention
    # Mask prevents looking at "future" dimensions
    mask = Upper_Triangular_Mask(size = output_dim + 1)
    context_vectors = Transformer_Encoder(full_sequence, mask)
    
    # 5. Output Heads
    # Map back to discrete bins
    logits = Linear_Head(context_vectors[offset_by_1])
    Return Logits
\end{lstlisting}

Similarly, the code for a RNN/GRU based solution is given by
\begin{lstlisting}
# Architecture: GRU-based Autoregressive Binner
Procedure JonasNet_RNN(x_input, y_targets, schedule_sampling_mask):
    # 1. Initial State
    x_feat = Feature_Extractor(x_input)
    hidden_state = Linear_Map(x_feat)  # h_0
    
    # 2. Sequential Decoding
    current_val = Zero_Tensor
    all_logits = []
    
    For d in 0 to output_dim - 1:
        # Recurrent Update
        output, hidden_state = GRU_Step(current_val, hidden_state)
        
        # Predict Bins
        logits_d = Bin_Head(output)
        all_logits.append(logits_d)
        
        # 3. Transition Logic (Input for next step)
        If Training and Scheduled_Sampling:
            # Mix ground truth and model prediction
            pred_val = ArgMax(Softmax(logits_d)) / n_bins
            current_val = Mix(y_targets[d], pred_val, schedule_sampling_mask)
        Else:
            # Inference: use previous prediction
            current_val = ArgMax(Softmax(logits_d)) / n_bins
            
    Return Stack(all_logits)
\end{lstlisting}

The code for reproducing the low-rank version for independent random variables $Y_1, \dots Y_n$ is
\begin{lstlisting}
# Architecture: Low-Rank Matrix Factorization (LoRA-style)
Procedure JonasNet_LoRA_Independent(x_input):
    # 1. Latent Encoding
    h = Feature_Extractor(x_input)
    
    # 2. Low-Rank Projection (Bottleneck)
    # Project to a smaller space to find shared structures
    z = SiLU(Linear_A(h))
    z = z + SiLU(Linear_B(z))  # Residual-like refinement
    z_compressed = SiLU(Linear_C(z))
    
    # 3. Reconstruction to Distribution Space
    # Expand compressed features to (output_dim * n_bins)
    flat_logits = Linear_D(z_compressed)
    
    # 4. Reshape & Normalize
    # Structure: [Batch, Dimension, Bins]
    grid_logits = Reshape(flat_logits, shape=(output_dim, n_bins))
    Return Log_Softmax(grid_logits, across_bins)
\end{lstlisting}

Sampling from the discretized probability mass function can be achieved via the multinomial distributions:
\begin{lstlisting}
Procedure Predict_Samples(X_raw, n_samples):
    # 1. Input Expansion
    # Expand input for parallel Monte Carlo paths: 
    # [Batch, Feat] -> [Batch * n_samples, Feat]
    X_expanded = Repeat_Interleave(X_raw, n_samples)
    total_batch = Length(X_expanded)
    
    # 2. Latent Feature Extraction
    # Map raw features to the model's latent dimension
    x_token = Feature_Extractor(X_expanded).Add_Dimension(1)
    
    # Initialize containers
    # current_y_seq stores the trajectory of sampled values for the Transformer
    current_y_seq = Initialize_Zeros(total_batch, output_dim, 1)
    sampled_values = Initialize_Zeros(total_batch, output_dim)

    # 3. Autoregressive Monte Carlo Loop
    With No_Gradient_Calculation:
        For d from 0 to output_dim - 1:
            # Generate causal sequence embeddings
            y_emb = Embedding_Layer(current_y_seq)
            sequence = Concatenate([x_token, y_emb]) + Positional_Bias
            
            # Transformer Pass with Causal Masking
            mask = Causal_Mask(output_dim + 1)
            context = Transformer(sequence, mask)
            
            # 4. Stochastic Bin Selection
            # Extract logits for dimension 'd' and convert to PDF
            logits = Bin_Head(context[at_index_d_plus_1])
            probabilities = Softmax(logits)
            
            # Multinomial Sampling: Draw a bin index based on the weights
            bin_idx = Sample_From_Multinomial(probabilities)
            
            # 5. Continuous De-quantization (Jitter)
            # Find the physical boundaries of the sampled bin
            left_edge = bin_edges[bin_idx]
            right_edge = bin_edges[bin_idx + 1]
            
            # Sample uniformly within the bin to create a continuous value
            # val ~ U(left_edge, right_edge)
            val = left_edge + Random_Uniform(0, 1) * (right_edge - left_edge)
            
            # 6. Trajectory Update
            sampled_values[:, d] = val
            If d < output_dim - 1:
                # Feed this sampled value back in for the next dimension d+1
                current_y_seq[:, d + 1] = val

    # 7. Final Tensor Reshaping
    # Fold the expanded batch back into (Batch, Samples, Dimensions)
    Return Reshape(sampled_values, (original_batch_size, n_samples, output_dim))
\end{lstlisting}

\bibliography{paper}

\begin{thebibliography}{6}
\providecommand{\natexlab}[1]{#1}
\providecommand{\url}[1]{\texttt{#1}}
\expandafter\ifx\csname urlstyle\endcsname\relax
  \providecommand{\doi}[1]{doi: #1}\else
  \providecommand{\doi}{doi: \begingroup \urlstyle{rm}\Url}\fi

\bibitem[Hollmann et~al.(2022)Hollmann, M{\"u}ller, Eggensperger, and Hutter]{hollmann2022tabpfn}
Noah Hollmann, Samuel M{\"u}ller, Katharina Eggensperger, and Frank Hutter.
\newblock Tabpfn: A transformer that solves small tabular classification problems in a second.
\newblock \emph{arXiv preprint arXiv:2207.01848}, 2022.

\bibitem[Kneib et~al.(2023)Kneib, Silbersdorff, and S{\"a}fken]{kneib2023rage}
Thomas Kneib, Alexander Silbersdorff, and Benjamin S{\"a}fken.
\newblock Rage against the mean--a review of distributional regression approaches.
\newblock \emph{Econometrics and Statistics}, 26:\penalty0 99--123, 2023.

\bibitem[Gneiting and Katzfuss(2014)]{gneiting2014probabilistic}
Tilmann Gneiting and Matthias Katzfuss.
\newblock Probabilistic forecasting.
\newblock \emph{Annual Review of Statistics and Its Application}, 1\penalty0 (1):\penalty0 125--151, 2014.

\bibitem[Gneiting(2008)]{gneiting2008probabilistic}
Tilmann Gneiting.
\newblock Probabilistic forecasting, 2008.

\bibitem[Gneiting and Raftery(2007)]{gneiting2007strictly}
Tilmann Gneiting and Adrian~E Raftery.
\newblock Strictly proper scoring rules, prediction, and estimation.
\newblock \emph{Journal of the American statistical Association}, 102\penalty0 (477):\penalty0 359--378, 2007.

\bibitem[Gneiting(2011)]{gneiting2011making}
Tilmann Gneiting.
\newblock Making and evaluating point forecasts.
\newblock \emph{Journal of the American Statistical Association}, 106\penalty0 (494):\penalty0 746--762, 2011.

\end{thebibliography}

\end{document}